\documentclass[12pt]{article}

\usepackage{graphicx}
\begin{document}

\begin{center}
{\bf Magnetic black holes within Einstein-AdS gravity coupled to nonlinear electrodynamics, extended phase space thermodynamics and Joule--Thomson expansion} \\
\vspace{5mm} S. I. Kruglov
\footnote{E-mail: serguei.krouglov@utoronto.ca}
\underline{}
\vspace{3mm}

\textit{Department of Physics, University of Toronto, \\60 St. Georges St.,
Toronto, ON M5S 1A7, Canada\\
Canadian Quantum Research Center, \\
204-3002 32 Ave Vernon, BC V1T 2L7, Canada} \\
\vspace{5mm}
\end{center}
\begin{abstract}

Einstein's gravity in AdS space coupled to nonlinear electrodynamics is studied. We analyse the metric, mass functions and corrections to the Reissner--Nordstr\"{o}m solution. Magnetic black holes thermodynamics in extended phase space is investigated. We formulate
the first law of  black hole thermodynamics showing that the generalized Smarr relation holds. The black hole stability is studied
by evaluating the Gibbs free energy and heat capacity. To study the cooling and heating phase transitions of black holes we consider the Joule--Thomson isenthalpic expansion. The Joule--Thomson coefficient and the inversion temperature are calculated.

\end{abstract}


\section{Introduction}

Studying black holes could shed light on quantum gravity which is not developed yet. In the last decades the behavior of black holes is a subject of intensive investigations which was stimulated recently by the  Event Horizon Telescope collaboration, detected the shadows of M87*  and   Sagittarius A* black holes \cite{EHT}. Hawking and Bekenstein showed that black holes are the thermodynamic systems \cite{Hawking,Bekenstein,Bekenstein1} (see also \cite{Bardeen,Jacobson,Padmanabhan}) with the entropy and the temperature connected with black holes surface area and surface gravity, respectively. The four laws of black hole mechanics were determined which are analogous to the ordinary thermodynamics
laws \cite{Bardeen}. The interest in Anti-de Sitter (AdS) space-time is due to AdS/CFT correspondence (the holographic principle) \cite{Maldacena} showing the connection of gravity in AdS space-time with the conformal field theory (CFT). The holographic principle has applications in condensed matter physics. Hawking and Page discovered that black holes first-order phase transitions take place in AdS space-time with negative cosmological constant \cite{Page}. Within AdS/CFT correspondence such phase transitions correspond to a confinement/deconfinement phase transitions \cite{Witten}. It was shown in \cite{Chamblin,Chamblin1,Rajagopal,Delsate,Pourhassan,Xu} that phase transitions in AdS black holes mimic Van der Waals liquid-gas phase transitions. Einstein-AdS black hole thermodynamics in an extended phase space was studied \cite{Dolan,Kubiznak,Mann,Teo}, where the cosmological constant is treated as thermodynamic pressure and the black hole mass is interpreted as enthalpy.

Black hole thermodynamics in Einstein-AdS gravity coupled to Born--Infeld electrodynamics, which is a special case of nonlinear electrodynamics (NED), was studied in \cite{Fernando,Dey,Cai,Fernando1,Myung,Banerjee,Miskovic,Chemissany,Balart,Meng,Falciano,Pourhassan1,Pourhassan2} and in extended phase space in \cite{Mann1,Zou,Hendi,Hendi1,Zeng}. Black hole thermodynamics in other NED models coupled to Einstein-AdS gravity was investigated in \cite{Kruglov,Kruglov1,Kruglov2,Kruglov3,Kruglov0}. The Joule–Thomson black
hole expansion firstly was studied in  \cite{Aydiner,Aydiner1} and later in
 \cite{Yaraie,Rizwan,Chabab,Mirza,Kruglov,Kruglov1,Kruglov2,Kruglov3,Kruglov4}.

 The interest in the model under consideration is in its simplicity. The mass and metric functions are expressed via elementary functions. The studying different models is useful to see different details in its behavior.

The structure of paper is as follows. In Section 2 we find and analyse the metric function possessing asymptotic with corrections to the Reissner--Nordstr\"{o}m solution. The first law of black hole thermodynamics in the extended phase space is formulated in Section 3. The thermodynamic magnetic potential and the conjugate to the NED coupling are calculated. It is proven that the generalized Smarr relation holds. In Section 4 we study the stability of black holes. The critical temperature, critical pressure, the Gibbs free energy and heat capacity are calculated. It is shown that phase transitions occur and black hole thermodynamics is similar to Van der Waals thermodynamics. In Section 5 the Joule--Thomson adiabatic expansion is investigated. We calculate the Joule--Thomson coefficient and the inversion temperature. In Section 6 we make a conclusion.

Throughout the paper the units $c=\hbar=1$, $k_B=1$ are used.

\section{Black hole solution of Einstein-AdS gravity coupled to NED}

The action of Einstein-AdS gravity coupled to NED is given by \cite{Fernando}
\begin{equation}
I=\int d^{4}x\sqrt{-g}\left(\frac{R-2\Lambda}{16\pi G_N}+\mathcal{L}(\mathcal{F}) \right),
\label{1}
\end{equation}
where $G_N$ is the Newton's constant, $\Lambda=-3/l^2$ being the negative cosmological constant and $l$ is the AdS radius. We consider the NED Lagrangian in the form \cite{Kruglov5}
\begin{equation}
{\cal L}(\mathcal{F}) =-\frac{{\cal F}}{4\pi\left(1+(2\beta{\cal F})^{3/2}\right)},
\label{2}
\end{equation}
with the Lorentz invariant ${\cal F}=F^{\mu\nu}F_{\mu\nu}/4=(B^2-E^2)/2$, and $E$ and $B$ are the electric and magnetic fields, correspondingly. At the limit $\beta\rightarrow 0$ Lagrangian (2) is converted into the Maxwell Lagrangian.
The field equations obtained from action (1) are
\begin{equation}
R_{\mu\nu}-\frac{1}{2}g_{\mu \nu}R+\Lambda g_{\mu \nu} =8\pi G_N T_{\mu \nu},
\label{3}
 \end{equation}
\begin{equation}
\partial _{\mu }\left( \sqrt{-g}\mathcal{L}_{\mathcal{F}}F^{\mu \nu}\right)=0,
\label{4}
\end{equation}
where $\mathcal{L}_{\mathcal{F}}=\partial \mathcal{L}( \mathcal{F})/\partial \mathcal{F}$. The energy-momentum tensor is given by
\begin{equation}
 T_{\mu\nu }=F_{\mu\rho }F_{\nu }^{~\rho }\mathcal{L}_{\mathcal{F}}+g_{\mu \nu }\mathcal{L}\left( \mathcal{F}\right).
\label{5}
\end{equation}
We explore the line element with the spherical symmetry
\begin{equation}
ds^{2}=-f(r)dt^{2}+\frac{1}{f(r)}dr^{2}+r^{2}\left( d\theta
^{2}+\sin ^{2}(\theta) d\phi ^{2}\right).
\label{6}
\end{equation}
The black hole is considered here as a magnetic monopole where the magnetic field is $B=q/r^2$ and $q$ is the magnetic charge.
The metric function found from Eq. (3) is \cite{Bronnikov}
\begin{equation}
f(r)=1-\frac{2m(r)G_N}{r},
\label{7}
\end{equation}
where the mass function is defined as
\begin{equation}
m(r)=m_0+4\pi\int_{0}^{r}\rho (r)r^{2}dr,
\label{8}
\end{equation}
with $m_0$ being the Schwarzschild mass (an integration constant), and $\rho$ is the energy density.
We obtain from Eq. (5) the magnetic energy density including the AdS space-time energy density
\begin{equation}
\rho=\frac{q^2r^2}{8\pi\left(r^6+\beta^{3/2}q^3\right)}-\frac{3}{8\pi G_Nl^2}.
\label{9}
\end{equation}
Making use of Eqs. (8) and (9) one finds the mass function
\[
m(r)=m_0+\frac{q^{3/2}}{24\sqrt[4]{\beta }}\biggl[
\sqrt{3}\ln\frac{r^2-\sqrt{3}\sqrt[4]{\beta q^2}r+\sqrt{\beta} q}{r^2+\sqrt{3}\sqrt[4]{\beta q^2}r+\sqrt{\beta} q}+4\arctan\left(\frac{r}{\sqrt[4]{\beta q^2}}\right)
\]
\begin{equation}
+2\arctan\left(\sqrt{3}+\frac{2r}{\sqrt[4]{\beta q^2}}\right)
-2\arctan\left(\sqrt{3}-\frac{2r}{\sqrt[4]{\beta q^2}}\right)\biggr]-\frac{r^3}{2G_Nl^2}.
\label{10}
\end{equation}
We obtain the magnetic energy (magnetic mass) of the black hole
\begin{equation}
m_M=\frac{q^2}{2}\int_0^\infty \frac{r^4}{r^6+(\beta q^2)^{3/2}}dr=\frac{\pi q^{3/2}}{6\sqrt[4]{\beta}},
\label{11}
\end{equation}
which is finite. As a result, the NED leads to the absence of singularities. By virtue of Eqs. (7) and (10) one finds the metric function
\[
f(r)=1-\frac{2m_0 G_N}{r}-\frac{q^{3/2}G_N g(r)}{12\sqrt[4]{\beta}r}+\frac{r^2}{l^2},
\]
\[
g(r)=\sqrt{3}\ln\frac{r^2-\sqrt{3}\sqrt[4]{\beta q^2}r+\sqrt{\beta} q}{r^2+\sqrt{3}\sqrt[4]{\beta q^2}r+\sqrt{\beta} q}+4\arctan\left(\frac{r}{\sqrt[4]{\beta q^2}}\right)
\]
\begin{equation}
+2\arctan\left(\sqrt{3}+\frac{2r}{\sqrt[4]{\beta q^2}}\right)
-2\arctan\left(\sqrt{3}-\frac{2r}{\sqrt[4]{\beta q^2}}\right).
\label{12}
\end{equation}
In the case when the Schwarzschild mass is zero ($m_0=0$) and as $r\rightarrow 0$, we obtain
\begin{equation}
f(r)=1+\frac{r^2}{l^2}-\frac{G_Nr^4}{5q\beta^{3/2}}+\frac{G_Nr^{10}}{11q^4\beta^{3}}+{\cal O}(r^{13}).
\label{13}
\end{equation}
Thus, the black hole is regular ($f(0)=1)$ and the metric function (13) has a de-Sitter core.
With the help of Eq. (12), when $\Lambda=0$, and as $r\rightarrow \infty$ one finds
\begin{equation}
f(r)=1-\frac{2MG_N}{r}+\frac{q^2G_N}{r^2}+\mathcal{O}(r^{-3}).
\label{14}
\end{equation}
As a result, $M=m_0+m_M$ is the ADM mass. Equation (14) shows that black holes possess corrections to the Reissner--Nordstr\"{o}m solution. At $\beta=0$ the metric (14) is converted into the Reissner--Nordstr\"{o}m metric. The metric function (12) plots are given in Fig. 1 at $m_0=0$, $G_N=1$, $l=15$.
\begin{figure}[h]
\includegraphics {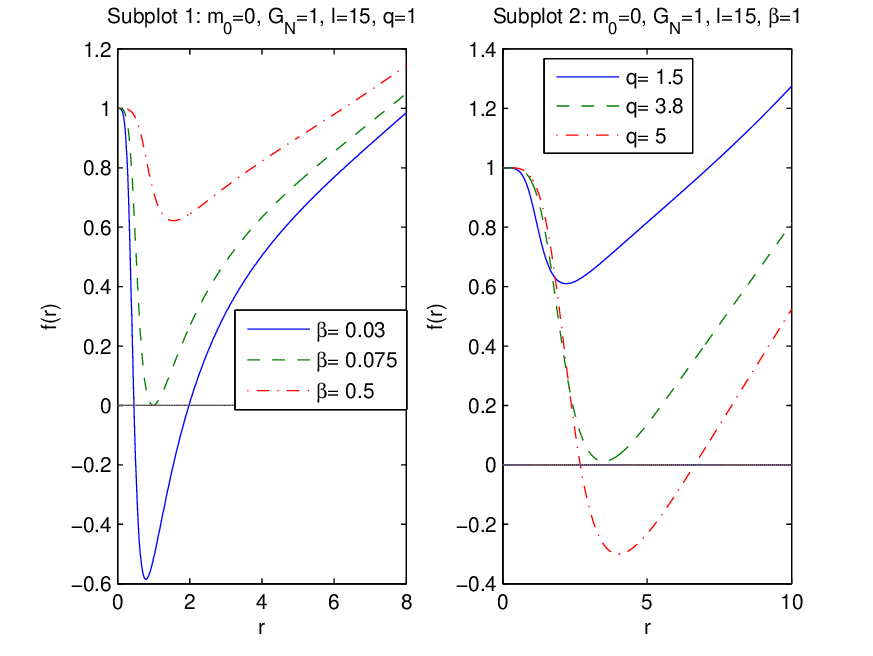}
\caption{\label{fig.1} The metric function $f(r)$ as a function of radius $r$ at $m_0=0$, $G_N=1$, $l=15$. The left panel of Fig. 1 shows that when coupling $\beta$ increases the event horizon radius decreases. In accordance with right panel of Fig. 1 if magnetic charge $q$ increases the event horizon radius also increases.}
\end{figure}
Figure 1 shows that black holes can have one or two horizons. If coupling $\beta$ increases the event horizon radius decreases. When magnetic charge $q$ increases, the event horizon radius also increases.

\section{First law of black hole thermodynamics and Smarr relation}

In extended phase space the positive pressure is given by $P=-\Lambda/(8\pi)$  \cite{Kastor,Dolan1,Cvetic,Kubiznak1,Kubiznak2} and coupling $\beta$ is a thermodynamic variable which is conjugated to vacuum polarisation. The black hole mass $M$ is treated as a chemical enthalpy ($M=U+PV$, $U$ is the internal energy). With the help of the Euler's dimensional analysis with $G_N=1$  \cite{Smarr}, \cite{Kastor}
we have dimensions as $[M]=L$, $[S]=L^2$, $[P]=L^{-2}$, $[J]=L^2$, $[q]=L$, $[\beta]=L^2$ and
\begin{equation}
M=2S\frac{\partial M}{\partial S}-2P\frac{\partial M}{\partial P}+2J\frac{\partial M}{\partial J}+q\frac{\partial M}{\partial q}+2\beta\frac{\partial M}{\partial \beta}.
\label{15}
\end{equation}
The vacuum polarization is ${\cal B}=\partial M/\partial \beta $ \cite{Teo}. For non-rotating black holes one has $J=0$;
the black hole entropy $S$, volume $V$ and pressure $P$ are given by
\begin{equation}
S=\pi r_+^2,~~~V=\frac{4}{3}\pi r_+^3,~~~P=-\frac{\Lambda}{8\pi}=\frac{3}{8\pi l^2}.
\label{16}
\end{equation}
From Eq. (12) we find the black hole mass that is understood as a chemical enthalpy
\begin{equation}
M(r_+)=\frac{r_+}{2G_N}+\frac{r_+^3}{2G_Nl^2}+\frac{\pi q^{3/2}}{6\sqrt[4]{\beta}}-\frac{q^{3/2}g(r_+)}{24\sqrt[4]{\beta}},
\label{17}
\end{equation}
with $r_+$ being the event horizon radius, $f(r_+)=0$.
Making use of Eq. (17) one obtains the vacuum polarization $\textit{B}$ and the thermodynamic potential $\Phi$
\[
\textit{B}=\frac{\partial M(r_+)}{\partial \beta} =\frac{q^{3/2}g(r_+)}{96\beta^{5/4}}-\frac{\pi q^{3/2}}{24\beta^{5/4}}+\frac{q^2r_+^5}{8\beta(r_+^6+(q\sqrt{\beta})^3)},
\]
\begin{equation}
\Phi=\frac{\partial M(r_+)}{\partial q}=\frac{\pi\sqrt{q}}{4\sqrt[4]{\beta}}-\frac{\sqrt{q}g(r_+)}{16\sqrt[4]{\beta}}+
\frac{qr_+^5}{4(r_+^6+(q\sqrt{\beta})^3)}.
\label{18}
\end{equation}
The Hawking temperature is defined as
\begin{equation}
T=\frac{f'(r)|_{r=r_+}}{4\pi},
\label{19}
\end{equation}
where $f'(r)=\partial f(r)/\partial r$. By virtue of Eq. (17) we obtain
\begin{equation}
\frac{\partial M(r_+)}{\partial r_+}=\frac{1}{2}+\frac{3r_+^2}{2l^2}-\frac{q^2r_+^4}{2(r_+^6+(q\sqrt{\beta})^3)}.
\label{20}
\end{equation}
 From Eqs. (12), (19) and (20), one finds the Hawking temperature
\begin{equation}
T=\frac{1}{4\pi}\left(\frac{1}{r_+}+\frac{3r_+}{l^2}-\frac{q^2r_+^3}{r_+^6+(q\sqrt{\beta})^3}\right).
\label{21}
\end{equation}
As $\beta\rightarrow 0$ Eq. (21) becomes the Hawking temperature of Maxwell-AdS black hole.
With the aid of Eqs. (16), (18) and (21) we obtain the first law of black hole thermodynamics
\begin{equation}
dM = TdS + VdP + \Phi dq + {\cal B}d\beta.
\label{22}
\end{equation}
Equation (22) represents the modification of ordinary the first law of thermodynamics by the additional term ${\cal B}d\beta$.
The plots of $\Phi$ and ${\cal B}$ versus $r_+$ are given in Fig. 2.
\begin{figure}[h]
\includegraphics {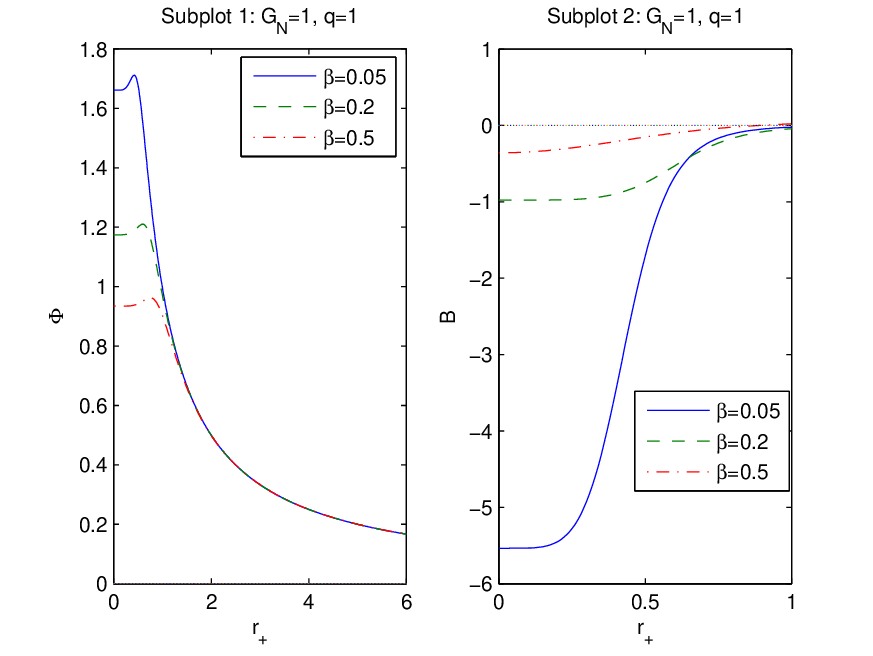}
\caption{\label{fig.2} The functions $\Phi$ and ${\cal B}$ versus $r_+$ at $q=1$. The solid curve in the left panel corresponds to $\beta=0.05$, the dashed curve is for $\beta=0.2$, and the dashed-doted curve corresponds to $\beta=0.5$. One can see that the magnetic potential $\Phi$ is finite at $r_+=0$ and is zero as $r_+\rightarrow \infty$. When NED parameter $\beta$ increases the magnetic potential decreases. The function ${\cal B}$ in the right panel becomes zero as $r_+\rightarrow \infty$ and is finite at $r_+=0$. }
\end{figure}
In accordance with Fig. 2 (left panel) when parameter $\beta$ increases the magnetic potential $\Phi$ decreases. When $r_+\rightarrow \infty$ the magnetic potential vanishes ($\Phi(\infty)=0$), and at $r_+ = 0$ potential $\Phi$ is finite. Making use of Eq. (18) one finds $\Phi(0)=\pi\sqrt{q}/(4\sqrt[4]{\beta})$ ($g(0)=0$). According to Fig. 2 there is a maxima of the functions $\Phi(r_+)$. When $\beta\rightarrow 0$ we have $\Phi\rightarrow q/(4r_+)$.
Figure 2 (right panel) shows that when $r_+ = 0$ the vacuum polarization is finite and as $r_+\rightarrow \infty$, vacuum polarisation ${\cal B}$ vanishes (${\cal B}(\infty)=0$).

From Eqs. (15), (16), (17) and (22) we obtain the generalized Smarr relation
\begin{equation}
M=2ST-2PV+q\Phi+2\beta{\cal B}.
\label{23}
\end{equation}

\section{Thermodynamics of black holes and phase transitions}

Making use of Eq. (21) we obtain the black hole equation of state (EoS)
\begin{equation}
P=\frac{T}{2r_+}-\frac{1}{8\pi r_+^2}+\frac{q^2r_+^2}{8\pi(r_+^6+(q\sqrt{\beta})^3)}.
\label{24}
\end{equation}
As $\beta\rightarrow 0$ Eq. (24) becomes EoS of charged  Maxwell-AdS black hole \cite{Kubiznak1}.
Equation (24) mimics the Van der Waals EoS if the specific volume is treated as $v=2l_Pr_+$ ($l_P=\sqrt{G_N}=1$) \cite{Kubiznak1}. Then Eq. (24) becomes
\begin{equation}
P=\frac{T}{v}-\frac{1}{2\pi v^2}+\frac{2q^2v^2}{\pi (v^6+64(q\sqrt{\beta})^3)}.
\label{25}
\end{equation}
Critical points (inflection points) of the $P-v$ diagrams obey equations as follows:
\[
\frac{\partial P}{\partial v}=-\frac{T}{v^2}+\frac{1}{\pi v^3} +\frac{8q^2v(32(q\sqrt{\beta})^3-v^6)}{\pi(v^6+64(q\sqrt{\beta})^3)^2}=0,
\]
\begin{equation}
\frac{\partial^2 P}{\partial v^2}=\frac{2T}{v^3}-\frac{3}{\pi v^4}+\frac{8q^2(5v^{12}-800(q\sqrt{\beta} )^3v^6+2048(q\sqrt{\beta} )^6)}{\pi(v^6+64(q\sqrt{\beta} )^3)^3}=0.
\label{26}
\end{equation}
With the help of Eq. (26) we obtain the equation for critical points $v_c$
\begin{equation}
(v_c^6+64(q\sqrt{\beta})^3)^3
-8q^2v_c^4(3v_c^{12}-864(q\sqrt{\beta})^3v_c^6+6144(q\sqrt{\beta})^6)=0.
\label{27}
\end{equation}
The critical temperature and pressure follow from  Eqs. (25) and (26)
\begin{equation}
T_c=\frac{1}{\pi v_c}+\frac{8q^2v_c^3(32(q\sqrt{\beta})^3-v_c^6)}{\pi(v_c^6+64(q\sqrt{\beta})^3)^2},
\label{28}
\end{equation}
\begin{equation}
P_c=\frac{1}{2\pi v_c^2}+\frac{6q^2v_c^2(64(q\sqrt{\beta})^3-v_c^6)}{\pi(v_c^6+64(q\sqrt{\beta})^3)^2}.
\label{29}
\end{equation}
There are two real solutions $v_c$ to Eq. (27) for each couplings $\beta$. In Table 1 we present only solutions for each $\beta$ that provide positive values of critical temperatures $T_{c}$ and pressures $P_{c}$.
\begin{table}[ht]
\caption{Critical values of the specific volume at $q=1$}
\centering
\begin{tabular}{c c c c c c c c c c c}\\[1ex]
\hline
$\beta$ & 0.1 & 0.2 & 0.3  & 0.4 & 0.5 & 0.6 & 0.7 & 0.8 & 0.9 & 1\\[0.5ex]
\hline
$v_{c}$ & 4.896 &4.891 & 4.885 & 4.877 & 4.868 & 4.857 & 4.846 & 4.833 & 4.819 & 4.804\\[0.5ex]
\hline
\end{tabular}
\end{table}
The $P-v$ diagrams are depicted in Fig. 3.
\begin{figure}[h]
\includegraphics {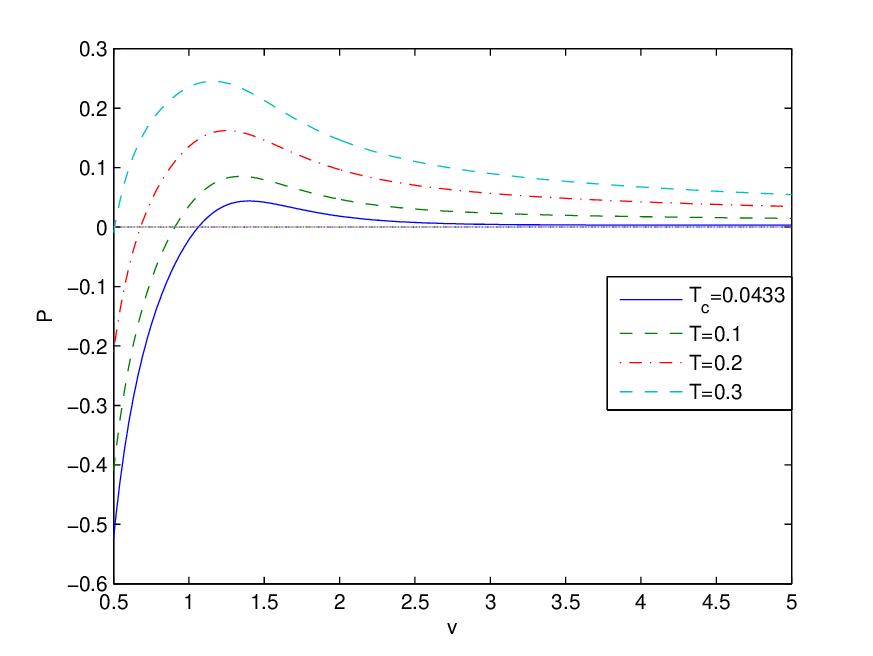}
\caption{\label{fig.3} The function $P(v)$ at $q=1$, $\beta=0.2$. The critical isotherm corresponds to $T_{c}\approx 0.0433$ having the inflection point.}
\end{figure}
At $q=1$, $\beta=0.2$ the critical specific volume is $v_{c}\approx 4.891$ and the critical temperature is $T_{c}\approx0.0433$ ($P_c\approx0.0033$).
According to Fig. 3 the pressure vanishes  at some points. When the specific volume increases (for small $v$) the pressure increases and the pressure possesses a maximum. After, the pressure decreases which is similar to the ideal gas. As $v\rightarrow 0$ the pressure $P\rightarrow -\infty$. This behavior is different from the Van der Waals liquid behavior at small $v$ (or $r$) where as $v\rightarrow 0$ the pressure $P\rightarrow \infty$. The coupling $\beta$ smoothing the singularity in third term (in right site) of Eq. (25) which becomes zero as $v\rightarrow 0$ and the dominant term is $-1/(2\pi v^2)$ which tends to $-\infty$. But at $\beta=0$ the dominant term is $2q^2/(\pi v^4)$ which goes to $+\infty$ as $v\rightarrow 0$ that is realised in Van der Waals liquid.
By virtue of Eqs. (27), (28) and (29) for small $\beta$, we obtain
\begin{equation}
v_c^2=24q^2+{\cal O}(\beta),~~~T_c=\frac{1}{3\sqrt{6}\pi q}+{\cal O}(\beta),~~~
P_c=\frac{1}{96\pi q^2}+{\cal O}(\beta).
\label{30}
\end{equation}
Equation (30) shows that at $\beta=0$ one obtains the critical points of charged AdS black hole \cite{Mann1}. The critical ratio becomes
\begin{equation}
\rho_c=\frac{P_cv_c}{T_c}=\frac{3}{8}+{\cal O}(\beta).
\label{31}
\end{equation}
The value $\rho_c=3/8$ corresponds to the Van der Waals fluid.

Because $M$ is understood as a chemical enthalpy, the Gibbs free energy (for fixed charge $q$, coupling $\beta$ and pressure $P$) is defined as
\begin{equation}
G=M-TS.
\label{32}
\end{equation}
From Eqs. (16), (17), (20) and (32) we find
\begin{equation}
G=\frac{r_+}{4}-\frac{2\pi r_+^3P}{3}+\frac{\pi q^{3/2}}{6\beta^{1/4}}+\frac{q^2r_+^5}{4(r_+^6+(q\sqrt{\beta})^3)}-\frac{q^{3/2}g(r_+)}{24\beta^{1/4}}.
\label{33}
\end{equation}
The plot of the Gibbs free energy $G$ versus $T$ for $\beta=0.2$ and $v_c\approx 4.891$, $T_c\approx 0.0433$ is given in Fig. 4. We used Eq. (24) showing that $P$ and $T$ are the functions of $r_+$.
\begin{figure}[h]
\includegraphics {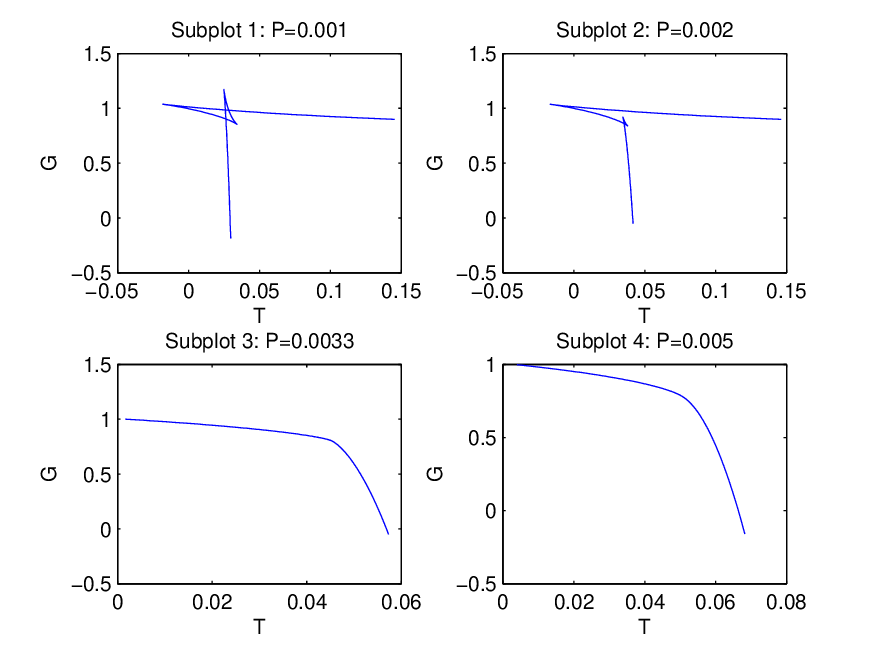}
\caption{\label{fig.4} The plots of the Gibbs free energy $G$ versus $T$ at $q=1$, $\beta=0.2$. Fig. 4 (subplots 1 and 2) shows 'swallowtail' plots resulting first-order phase transitions. Subplot 3 shows to second-order phase transition with $P=P_c\approx 0.0033$. Subplot 4 shows the case $P>P_c$ with non-critical behavior of the Gibbs free energy.}
\end{figure}
In accordance with subplots 1 and 2 in Fig. 4, at $P<P_c$ first-order phase transitions occur with the 'swallowtail' behaviour. Subplot 3 shows the second-order phase transition for $P=P_c$. According to subplot 4, at the case $P>P_c$ phase transitions are absent.

The entropy $S$ versus temperature $T$ at $q=1,~\beta=0.2$ is depicted in Fig. 5. Subplots 1 and 2 in Fig. 5 show ambiguous functions of entropy that show first-order phase transitions.  According to subplot 3 the second-order phase transition takes place.  Low and high entropy states are separated by the critical point. Subplot 4 shows that there are not second-order phase transitions at $q=1,~\beta=0.2$, $P=0.005$.
\begin{figure}[h]
\includegraphics {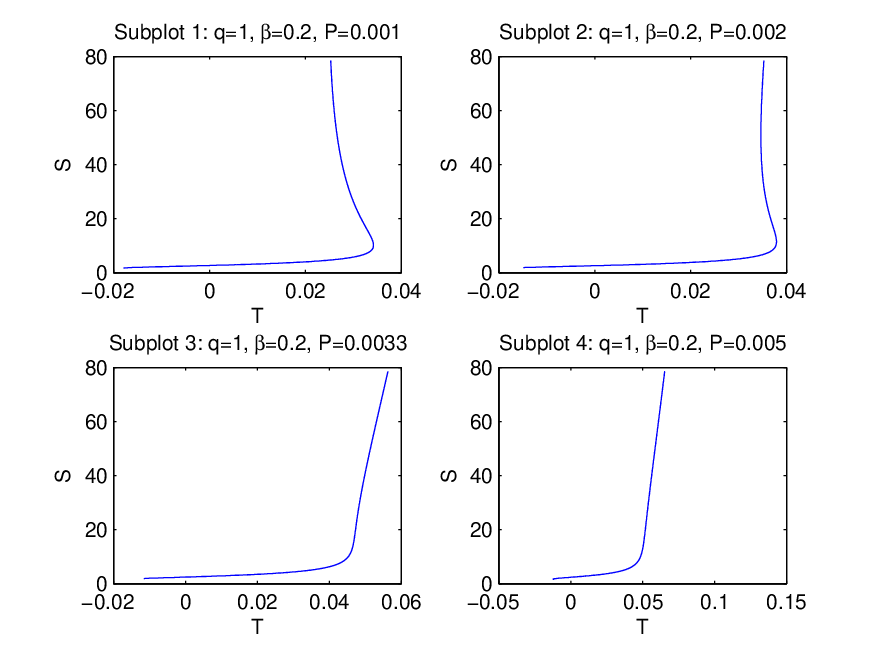}
\caption{\label{fig.5} The plots of entropy $S$ versus temperature $T$ at $q=1,~\beta=0.2$. In accordance with subplots 1 and 2 entropy is ambiguous function of the temperature and, therefore, first-order phase transitions take place. According to subplot 3 the second-order phase transition occurs.}
\end{figure}

\subsection{Heat capacity}

To study the black hole local stability we consider the heat capacity which is given by
\begin{equation}
C_q=T\left(\frac{\partial S}{\partial T}\right)_q=\frac{T\partial S/\partial r_+}{\partial T/\partial r_+}=\frac{2\pi r_+ T}{G_N\partial T/\partial r_+}.
\label{34}
\end{equation}
When the Hawking temperature has an extremum the heat capacity diverges and black hole phase transition occurs. The plot of the Hawking temperature is depicted in Fig. 6 for various parameters $\beta=0.1,~0.3,~1$ ($l=q=1$).
\begin{figure}[h]
\includegraphics {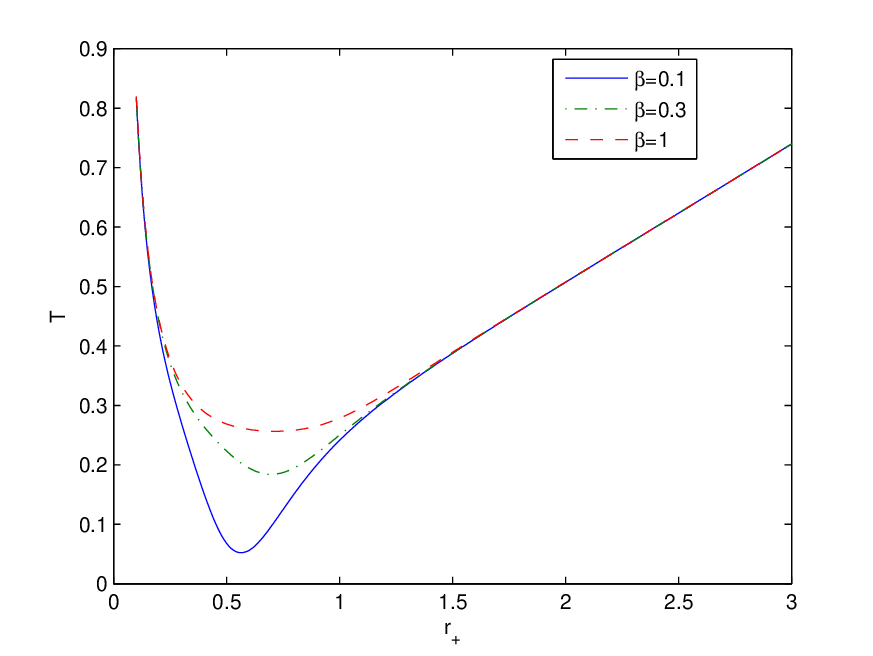}
\caption{\label{fig.6} The plots of the Hawking temperature $T$ versus horizon radius $r_+$ at $l=q=1,~\beta=0.1,~0.3,~1$. In accordance with figures, the Hawking temperature possesses minima.}
\end{figure}
According to Fig. 6, the Hawking temperature possesses minima, and therefore, the heat capacity has singularities. The plots of the heat capacity (34) at $q=l=1,~ \beta=0.1$ ($G_N=1$) is given in Fig. 7. Figure 7 shows that indeed the heat capacity diverges at the point where the Hawking temperature has a minimum.
\begin{figure}[h]
\includegraphics {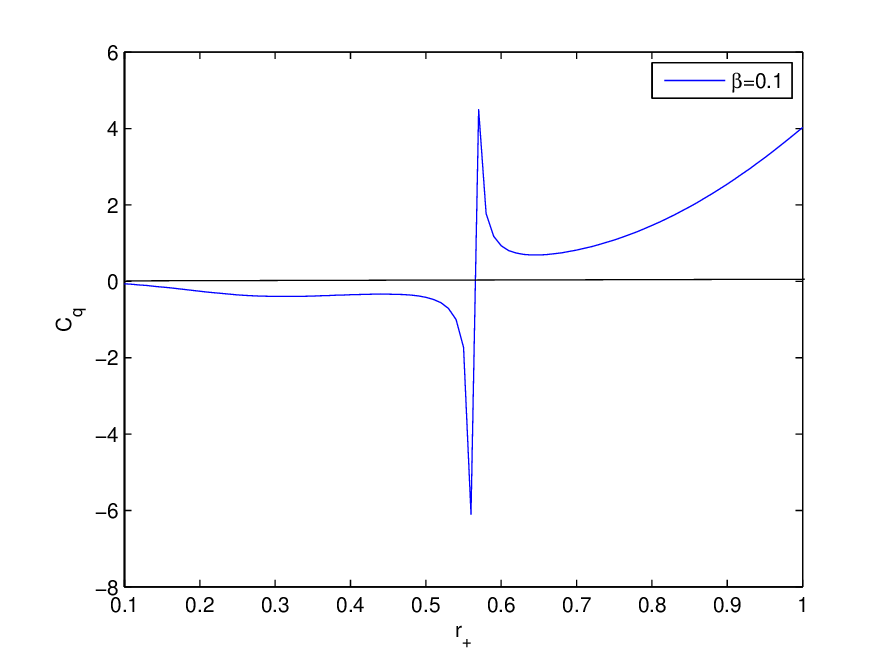}
\caption{\label{fig.7} The plot of the heat capacity $C_q$ versus horizon radius $r_+$ at $l=q=1,~\beta=0.1$. According to the figure, the heat capacity has a singularity where the Hawking temperature possesses a minimum.}
\end{figure}
In this point the phase transition occurs. In the region where the heat capacity is positive the black hole is locally stable, otherwise the black hole is unstable. The heat capacity possesses a singularity at $\partial T/\partial r_+$=0. Making use of Eq. (21) we find
\[
\frac{\partial T}{\partial r_+}=\frac{1}{4\pi}\left(-\frac{1}{r_+^2}+\frac{3}{l^2}-\frac{3q^2r_+^3((q\sqrt{\beta})^3-r_+^6)}{(r_+^6+(q\sqrt{\beta})^3)^2}\right).
\]
For the case $l=q=1,~\beta=0.1$ the approximate real and positive solution to equation $\partial T/\partial r_+=0$ is $r_s\approx 0.565$.
At this point the black hole undergoes the phase transition from small black hole to large black hole. According to Fig. 7 at
$r_+>r_s$ the black hole is stable but at $r_s>r_+>0$ the black hole is unstable. Thus, the small sized black holes are unstable and large sized black holes are stable.
One can verify that at any parameters we have $\mbox{lim}_{r_+\rightarrow 0}C_q=0$, i.e. the heat capacity vanishes at $r_+\rightarrow 0$. But at the infinitesimal radius quantum effects are
important \cite{Xavier,Behnam,Xavier1,Behnam1,Sudhaker,Ruben}. Therefore, without quantum corrections at small radius the stability analyses is not completed. Quantum gravity effects start at the Planck length $l_P=\sqrt{G_N\hbar/c^3}\approx1.6\times 10^{-35}$ m because of quantum fluctuations. It is worth mentioning that the smooth metric fields $g_{\mu\nu}$ of the classical gravity can not be used at $r\leq l_P$ but the idea of "space-time foam" \cite{Garlip} is fruitful for describing space-time microscopic degrees of freedom.

\section{The Joule--Thomson expansion and cooling-heating
phase transitions.}

The Joule--Thomson expansion takes place when the enthalpy, which is the black hole mass, is constant. This isenthalpic expansion is described by the Joule--Thomson coefficient
\begin{equation}
\mu_J=\left(\frac{\partial T}{\partial P}\right)_M=\frac{1}{C_P}\left[ T\left(\frac{\partial V}{\partial T}\right)_P-V\right]=\frac{(\partial T/\partial r_+)_M}{(\partial P/\partial r_+)_M},
\label{35}
\end{equation}
showing the cooling-heating phases. The heating phase takes place when $\mu_J<0$, and the cooling phase occurs if $\mu_J>0$.
The Joule--Thomson coefficient represents the slope of the $P-T$ function.
The sign of $\mu_J$ is changed at the inversion temperature $T_i$. The inversion temperature $T_i$ is defined by equation $\mu_J(T_i)=0$. When the initial temperature is higher than inversion temperature $T_i$ the cooling phase ($\mu_J>0$) occurs, and as a result,
the final temperature decreases. In the case when the initial temperature is lower than $T_i$, the final temperature increases corresponding to the heating phase ($\mu_J<0$). By taking Eq. (35) and  equation $\mu_J(T_i)=0$, one obtains
\begin{equation}
T_i=V\left(\frac{\partial T}{\partial V}\right)_P=\frac{r_+}{3}\left(\frac{\partial T}{\partial r_+}\right)_P.
\label{36}
\end{equation}Cooling and heating processes are separated by the inversion temperature. The line of inversion temperature
crosses $P-T$ diagrams maxima \cite{Yaraie,Rizwan}. Equation (24) can be written as EoS
\begin{equation}
T=\frac{1}{4\pi r_+}+2P r_+-\frac{q^2r_+^3}{4\pi(r_+^6+(q\sqrt{\beta})^3)}.
\label{37}
\end{equation}
When $\beta=0$ Eq. (37) becomes EoS of Maxwell-AdS black holes. Making use of Eq. (17) and equation $P=3/(8\pi l^2)$ we obtain
\begin{equation}
P=\frac{3}{4\pi r_+^3}\left[M(r_+)-\frac{r_+}{2}-\frac{\pi q^{3/2}}{6\beta^{1/4}}
+\frac{q^{3/2}g(r_+)}{24\beta^{1/4}}\right].
\label{38}
\end{equation}
The $P-T$ isenthalpic diagrams are depicted in Fig. 8 by using Eqs. (37) and (38).  The inversion $P_i-T_i$ diagram crosses maxima of isenthalpic curves in Fig. 8.
 \begin{figure}[h]
\includegraphics {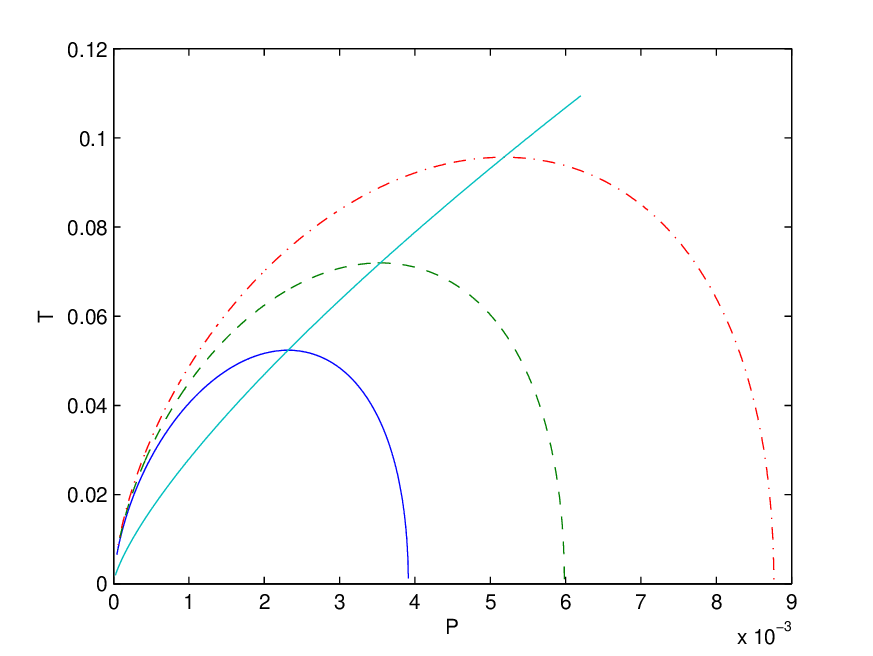}
\caption{\label{fig.8} The plots of the temperature $T$ versus pressure $P$ at $q=40$, $\beta=0.2$.
The $P_i-T_i$ diagram srosses maxima of isenthalpic curves. The solid curve corresponds to mass $M=100$, the dashed curve
is for $M=110$, and the dashed-doted curve corresponds to $M=120$. When black hole masses increase the inversion temperature $T_i$ also increases.}
\end{figure}
With the help of Eqs. (24), (36) and (37) one finds the inversion pressure $P_i$
\begin{equation}
P_i=\frac{3q^2r_+^8}{8\pi \left(r_+^6+(\beta q^2)^3\right)^2}-\frac{1}{4\pi r_+^2}.
\label{39}
\end{equation}
From Eqs. (37) and (39) one finds the inversion temperature
\begin{equation}
T_i=\frac{q^2r_+^3(2r_+^6-(q\sqrt{\beta})^3)}{4\pi(r_+^6+(q\sqrt{\beta})^3)^2}-\frac{1}{4\pi r_+}.
\label{40}
\end{equation}
Putting $P_i=0$ in Eq. (39), we obtain the equation for the minimum of the event horizon radius $r_{min}$
\begin{equation}
2(r_{min}^6+(\sqrt{\beta} q)^3)^2-3q^2r_{min}^{10}=0.
\label{41}
\end{equation}
Making use of Eqs. (40) and (41) at $\beta=0$, one comes to the minimum of the inversion temperature in accordance with Maxwell-AdS magnetic black holes
\begin{equation}
T_i^{min}=\frac{1}{6\sqrt{6}\pi q}, ~~~~r_{min}=\frac{\sqrt{6}q}{2}.
\label{42}
\end{equation}
From Eqs. (30) and (42) at $\beta=0$ we obtain the relation $T_i^{min}=T_c/2$ corresponding to Maxwell-AdS black holes \cite{Aydiner}.
By vurtue of Eqs. (39) and (40) $P_i-T_i$ diagrams are plotted in Fig. 8. Figure 8 shows that the inversion point increases when the black hole mass increases. The inversion diagrams $P_i-T_i$ are given in Figs. 9 and 10.
\begin{figure}[h]
\includegraphics {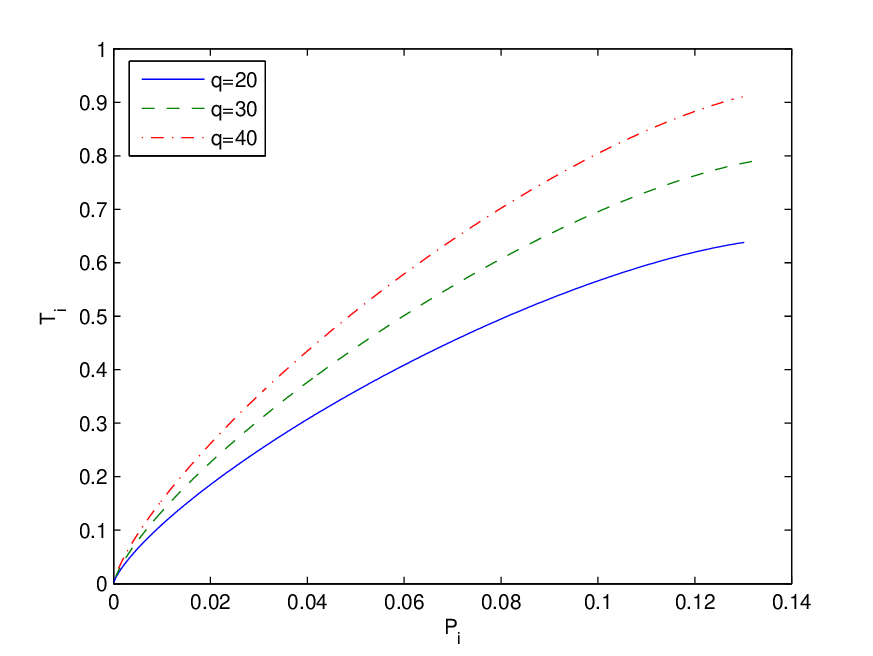}
\caption{\label{fig.9} The inversion temperature $T_i$ versus pressure $P_i$ at $q=20$, $30$ and $40$, $\beta=0.2$. When magnetic charge $q$ increases the inversion temperature also increases.}
\end{figure}
\begin{figure}[h]
\includegraphics {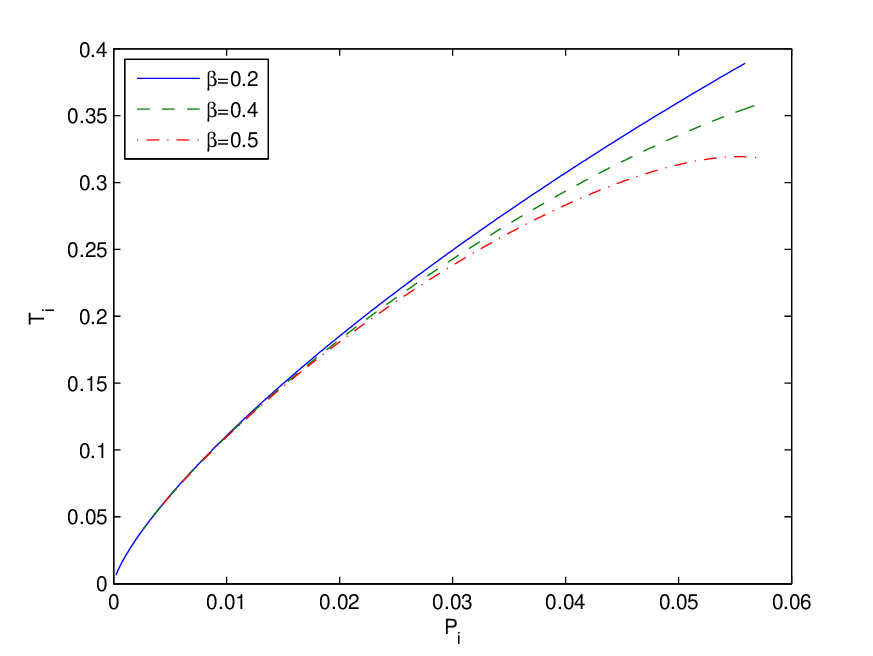}
\caption{\label{fig.10} The inversion temperature $T_i$ versus pressure $P_i$ at $\beta=0.2$, $0.4$ and $0.5$, $q=20$. When the coupling $\beta$ increases the inversion temperature decreases.}
\end{figure}
In accordance with Fig. 9 when magnetic charge $q$ increases, the inversion temperature also increases. According to
Fig. 10 when the coupling $\beta$ increases the inversion temperature decreases.
Making use of Eqs. (35), (37) and (39) we obtain
\[
\left(\frac{\partial T}{\partial r_+}\right)_M=-\frac{1}{4\pi r_+^2}+2P|_M+2r_+\left(\frac{\partial P}{\partial r_+}\right)_M
-\frac{3q^2r_+^2((q\sqrt{\beta})^3-r_+^6)}{4\pi(r_+^6+(q\sqrt{\beta})^3)^2},
\]
\begin{equation}
\left(\frac{\partial P}{\partial r_+}\right)_M=\frac{3}{4\pi r_+^4}\biggl[\frac{q^{3/2}\pi}{2\beta^{1/4}}-3M+r_+
-\frac{q^{3/2}g(r_+)}{8\beta^{1/4}}+\frac{q^2r_+^5}{2(r_+^6+(q\sqrt{\beta})^3)}\biggr],
\label{43}
\end{equation}
where $P|_M$ is given in Eq. (38).
From Eqs. (35) and (43) one can find the Joule--Thomson coefficient which is the function of the magnetic charge $q$, coupling $\beta$, black hole mass $M$ and event horizon radius $r_+$. In the case when the Joule--Thomson coefficient is positive, $\mu_J>0$, a cooling process takes place but when $\mu_J<0$ a heating process occurs.

\section{Conclusion}

New magnetic black hole solution in Einstein-AdS gravity coupled to NED has been obtained. We have found the metric, mass functions and showed that the magnetic mass of the black hole is finite. It was demonstrated that there are corrections to the Reissner--Nordstr\"{o}m solution (at $\Lambda=0$) due to NED parameter $\beta$. The total mass $M$ which is the sum of the Schwarzschild mass $m_0$ and  magnetic mass $m_M$ can be considered as ADM mass. We have found the asymptotic as $r\rightarrow 0$ (at $m_0=0$) with a de-Sitter core. Coupling $\beta$ smoothing singularity so that the black hole is regular. The plots of the metric function $f(r)$ shown that when coupling $\beta$ increases at constant magnetic charge the event horizon radius decreases and when magnetic charge increases at constant coupling $\beta$ the event horizon radius increases. We have formulated the first law of black hole thermodynamics in an extended phase space where positive thermodynamic pressure is connected with negative cosmological constant and dimensional coupling $\beta$ conjugates the vacuum polarisation. The thermodynamic potential and vacuum polarisation as functions of the event horizon radius have been found and we showed that the generalized Smarr relation holds.
We have studied the black holes thermodynamics and phase transitions where the mass of the black hole is treated as the chemical enthalpy.
Equation of state has shown that black hole thermodynamics is similar to the Van der Waals liquid–gas thermodynamics.
Critical points (inflection points) of the $P-v$ diagrams have been evaluated corresponding to second-order phase transitions.
We have demonstrated that the critical ratio $\rho_c$ depends in the coupling $\beta$ and it is different from the Van der Waals value $3/8$.
The Gibbs free energy has been analysed showing the first-order phase transitions with the 'swallowtail' behaviour and second-order phase transitions at critical points. The plots of the entropy $S$ versus temperature $T$ for various parameters have been depicted which is the
ambiguous function corresponding to first-order phase transitions. The heat capacity is calculated which possesses singularities. When he heat capacity is positive black holes are stable, otherwise they are unstable. We have studied the black hole Joule--Thomson isenthalpic expansions to analyse cooling-heating phase transitions. The Joule{Thomson coefficient and the inversion temperature separating cooling and heating processes of black holes have been found. The heating phase corresponds to $\mu_J < 0$, and the cooling phase takes place at $\mu_J > 0$. It was demonstrated that when the black hole mass increases the inversion point increases.

\vspace{3mm}
Competing interests: The authors declare there are no competing interests.
\vspace{3mm}

This manuscript does not report data.

\end{document}